\documentclass[a4paper,11pt]{article}
\usepackage{mathrsfs}
\usepackage{amssymb}
\usepackage{amsmath}
\usepackage{mathtools}
\usepackage{graphicx}
\usepackage{pos}

\usepackage{array}
\newcommand{\PreserveBackslash}[1]{\let\temp=\\#1\let\\=\temp}
\newcolumntype{C}[1]{>{\PreserveBackslash\centering}p{#1}}
\newcolumntype{R}[1]{>{\PreserveBackslash\raggedleft}p{#1}}
\newcolumntype{L}[1]{>{\PreserveBackslash\raggedright}p{#1}}

\newlength{\mycol}
\newcommand{\ud}{\, \mathrm{d}}
\newcommand{\cf}{C_\mathrm{F}}

\newcommand{\nc}{{N_\mathrm{c}}}

\newcommand{\der}{\mathrm{d}}
\newcommand{\rt}{{\mathbf{r}}}
\newcommand{\xt}{{\mathbf{x}}}
\newcommand{\xij}[1]{{\mathbf{x}_{#1}}}
\newcommand{\bt}{{\mathbf{b}}}

\newcommand{\zt}{{\mathbf{z}}}

\newcommand{\xbj}{{x_\text{Bj}}}
\newcommand{\Yobk}{{Y_{0, \text{BK}}}}
\newcommand{\Yoif}{{Y_{0, \text{if}}}}
\newcommand{\etaobk}{{\eta_{0, \text{BK}}}}

\newcommand{\sigmadip}{{ \sigma_\textrm{dip} }}

\newcommand{\sigmaltnlo}{\sigma_{L,T}^{\textrm{NLO}}}
\newcommand{\sigmaltdip}{\sigma_{L,T}^{\textrm{dip}}}
\newcommand{\sigmaltqg}{\sigma_{L,T}^{qg}}

\newcommand{\sigmaltqgu}{\sigma_{L,T}^{qg, \textrm{unsub.}}}
\newcommand{\sigmalt}[1]{\sigma_{L,T}^{#1}}

\newcommand{\zmin}{z_{2,\textrm{min}}}

\newcommand{\kcal}{\mathcal{K}}

\title{
Dipole model at Next-to-Leading Order meets HERA data
}

\author[a,b,c]{Guillaume Beuf}
\author*[a,b]{Henri Hänninen}
\author[a,b]{Tuomas Lappi}
\author[a,b]{Heikki Mäntysaari}

\affiliation[a]{
Department of Physics, University of Jyv\"askyl\"a, \\
 P.O. Box 35, 40014 University of Jyv\"askyl\"a, Finland
}

\affiliation[b]{
Helsinki Institute of Physics, \\ P.O. Box 64, 00014 University of Helsinki, Finland
}

\affiliation[c]{
National Centre for Nuclear Research, \\ 02-093, Warsaw, Poland
}

\emailAdd{guillaume.beuf@gmail.com}
\emailAdd{henri.j.hanninen@jyu.fi}
\emailAdd{tuomas.v.v.lappi@jyu.fi}
\emailAdd{heikki.mantysaari@jyu.fi}

\abstract{
Deep inelastic scattering (DIS) total cross section data at small-x as measured by the HERA experiments is well described by Balitsky-Kovchegov (BK) evolution in the leading order dipole picture. Recently the full Next-to-Leading Order (NLO) dipole picture total cross sections have become available for DIS, and a working factorization scheme has been devised which subtracts the soft gluon divergence present at NLO.
We report our recently published work in which we make the first comparisons of the NLO DIS total cross sections to HERA data. The non-perturbative initial condition to BK evolution is fixed by fitting the HERA reduced cross section data. As the NLO results for the DIS total cross section are currently available only in the massless quark limit, we also fit a light quark only--cross section constructed with a parametrization of published total and heavy quark data. We find an excellent description of the HERA data. Since the full NLO BK equation is computationally expensive, we use a number of beyond LO prescriptions for the evolution that include most important higher order corrections enhanced by large transverse logarithms, including the recent version of the equation formulated in terms of the target momentum fraction.
}

\FullConference{%
  HardProbes2020\\
  1-6 June 2020\\
  Austin, Texas}

\begin{document}
\maketitle

\section{Introduction}

In the Color Glass Condensate (CGC) framework deep inelastic scattering (DIS) proceeds as follows at leading order accuracy (LO). First, the incoming lepton can be factorized out and one is left with a virtual photon scattering from the color field of the proton. The virtual photon fluctuates into a quark-antiquark pair that then scatters eikonally from the color field. To get the total $\gamma^{*}p$ cross section one computes the forward scattering amplitude and applies the optical theorem. This procedure results in the cross sections at LO for the photon polarization states: 
\begin{equation}
\label{eq:lo-cross-section}
    \sigma^{\gamma^* p}_{T,L}=2 \int \der^2 \bt \der^2 \rt \der z |\psi^{\gamma^* \to q\bar q}(\rt,Q^2,z)|^2
    \left(1-S(\rt, \bt, x)\right).
\end{equation}
The scattering matrix $S$ is given as a correlator of the Wilson lines picked up by the quark and antiquark 
in the target.

The $\xbj$-evolution of the target color field is described by the JIMWLK equation, or approximatively by the Balitsky-Kovchegov (BK) equation~\cite{Balitsky:1995ub,Kovchegov:1999yj}, which is to leading order accuracy:
\begin{equation}
    \label{eq:bk-evolution}
    \frac{\partial S(\xij{01})}{\partial Y} =  \int \der^2 \xij{2} K_\text{BK}
    (\xij{0}, \xij{1}, \xij{2})
    [S(\xij{02}) S(\xij{21}) - S(\xij{01})].
\end{equation}

In this work~\cite{Beuf:2020dxl} we use three beyond leading order prescriptions of the BK equation that capture important higher order effects using resummation methods, which include in the equation higher order contributions enhanced by large transverse logarithms. The full next to leading order BK equation has also been solved numerically~\cite{Lappi:2016fmu}, however it is computationally quite expensive.

Two of the three versions of the BK equation use an evolution variable expressed in terms of the fraction of the projectile photon momentum, which leads to an evolution equation as a function of $Y \sim \ln W^2$.
This matches the convention in the computation of the DIS cross sections. The other option, studied recently in Ref.~\cite{Ducloue:2019ezk},  is to parametrize the evolution by the target momentum fraction so that the evolution variable is $\eta \sim \ln 1/\xbj$.

In the projectile momentum fraction parametrization the first of these enhanced BK equations is the Kinematically Constrained BK (KCBK) \cite{Beuf:2014uia}
		\begin{equation} \label{eq:kcbk}
			\partial_Y S(\xij{01},Y) =
				\! \int \! \der^2\zt K_{\text{BK}}
				\theta\left( Y \! - \! \Delta_{012} - \Yoif \right)
				\left[ S(\xij{02}, Y \! - \! \Delta_{012}) S(\xij{21}, Y \! - \! \Delta_{012}) - S(\xij{01}, Y)  \right].
		\end{equation}
Here the resummation procedure has lead to a non-local equation in the evolution variable.
The second is the Collinearly Resummed BK (ResumBK) \cite{Iancu:2015vea, Iancu:2015joa}
		\begin{equation} \label{eq:resumbk}
			\partial_Y S(\xij{01},Y) = \int \der^2 \xij{2} K_{\text{DLA}} K_{\text{STL}} K_{\text{BK}} [S(\xij{02}) S(\xij{21}) - S(\xij{01})],
		\end{equation}
which additionally incorporates a partial resummation of single transverse log enhanced contributions. This correction is separate from the first resummation discussed and in principle could be included in each of the BK equations discussed. We elected to work with established formulations of BK and so we did not add this correction to the other equations. However we did verify that the effects of this correction to the fits were minor.

For target momentum fraction evolution (TBK) we use the equation formulated in Ref.~\cite{Ducloue:2019ezk}:
	\begin{equation}
	\label{eq:trbk}
	\partial_\eta \bar S(\xij{01},\eta) = \int \der^2 \xij{2}  K_\text{BK} \theta(\eta-\eta_0 - \delta)
	[\bar S(\xij{02}, \eta - \delta_{02}) \bar S(\xij{21}, \eta - \delta_{21}) - \bar S(\xij{01}, \eta)  ].
	\end{equation}
\vspace{-0.1em}
This equation is non-local in the evolution variable as well. Also one must properly deal with the fact that the evolution momentum fraction $\sim e^{-\eta}$ does not directly  appear in the DIS cross section.

\section{Deep inelastic scattering in the dipole picture at next-to-leading order}

The next to leading order DIS structure functions have been derived in conventional dimensional regularization \cite{Beuf:2016wdz,Beuf:2017bpd}, and in the four dimensional helicity scheme \cite{Hanninen:2017ddy}. A working soft gluon factorization has been formulated \cite{Ducloue:2017ftk}, where the NLO cross sections were partitioned in a lowest order contribution and two NLO contributions:
\begin{equation}
    \sigmaltnlo = \sigmalt{\textrm{IC}} + \sigmaltqgu +\sigmaltdip.
\end{equation}
The two NLO corrections $\sigmaltqg$ and $\sigmadip$ 
roughly come from the tree and loop contributions to the photon wave functions, respectively. 
They can be written as (for explicit expressions see Ref.~\cite{Ducloue:2017ftk}):
\begin{align}
    \sigmaltqgu 
        &= 
        8 \nc \alpha_\text{em} \frac{\alpha_s \cf}{\pi} \sum_f e_f^2
        \int_0^1 \der z_1 \int_{\zmin}^{1-z_1} \frac{\der z_2}{z_2}
        \label{eq:NLO_qg_unsub}
        \int\displaylimits_{\xt_0, \xt_1, \xt_2}
        \mkern-4mu 
        \mathcal{K}_{L,T}^{\textrm{NLO}}(z_1, z_2, \xt_0, \xt_1, \xt_2),
\\
     \sigmaltdip
         &= 4 \nc \alpha_{em} \frac{\alpha_s \cf}{\pi} \sum_f e_f^2 \int_0^1 \ud z_1
        \label{eq:NLO_dip}
        \int\displaylimits_{\xt_0, \xt_1} 
        \mkern-9mu 
        \kcal_{L,T}^\text{LO}
        \!
        \left(z_1,\xt_0,\xt_1\right) 
        \! 
        \left[\frac{1}{2}\ln^2\!\left(\!\frac{z_1}{1\!-\!z_1}\!\right)\!-\!\frac{\pi^2}{6}\!+\!\frac{5}{2}\right].
\end{align}
The presence of the gluon in the term  $\sigmaltqg$ ties into the evolution length through the lower integration limit of the gluon fractional momentum:
    $ \zmin \equiv e^{\Yoif}  \frac{Q_0^2}{W^2}$.

\section{Fit results}

		\begin{figure}[t]
			\begin{minipage}[c]{0.48\textwidth}
				\includegraphics[width=\textwidth]{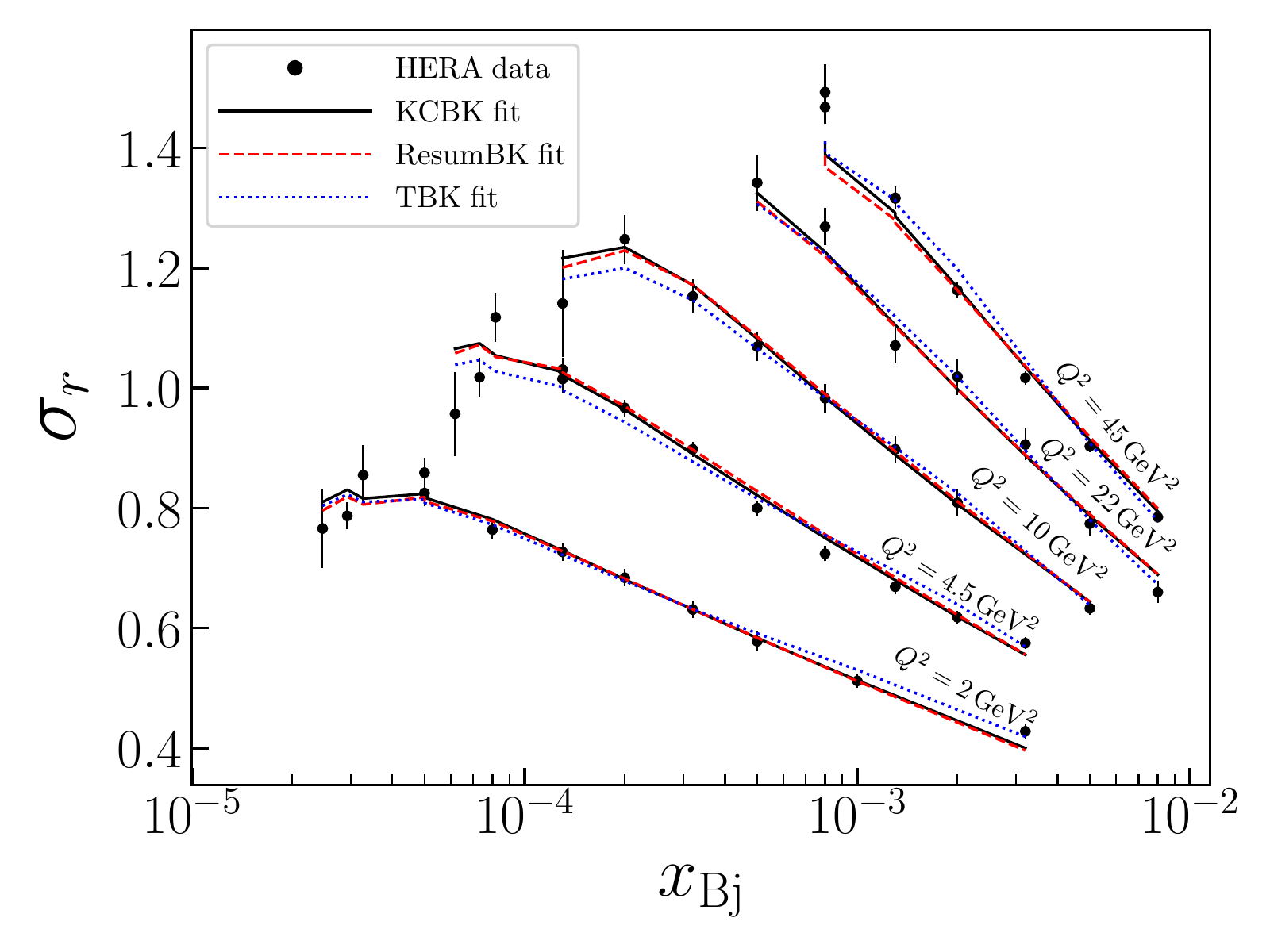}
				\caption{NLO fit to HERA data with each of the BK equations. Fits use Bal + SD coupling and $\Yobk = \etaobk = \ln 1/0.01$.}
				\label{fig:hera-sigmar}
			\end{minipage} 
				\hfill
			\begin{minipage}[c]{0.48\textwidth}
				\includegraphics[width=\textwidth]{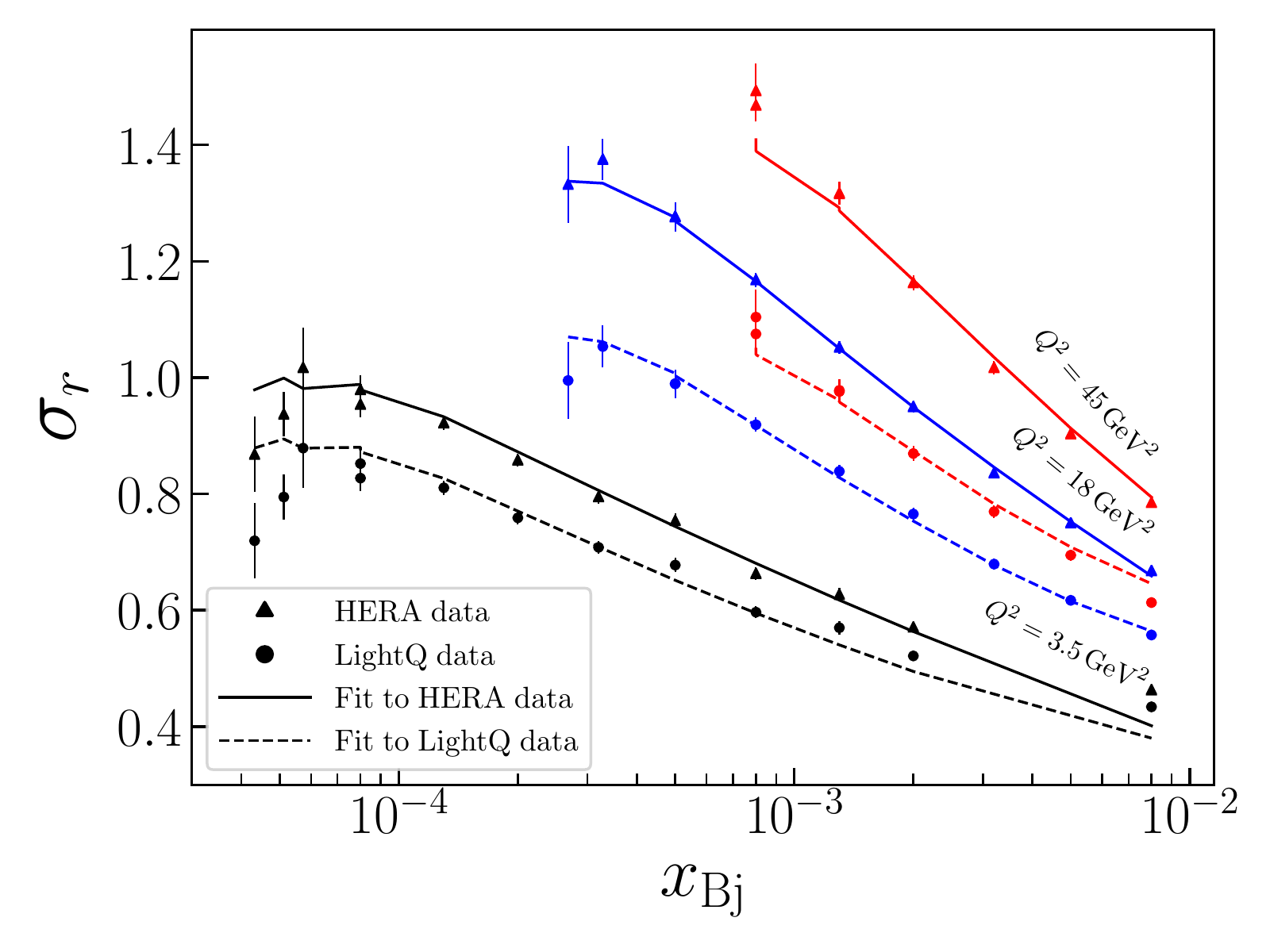}
				\caption{NLO fits to light quark data and HERA data using the KCBK with Bal + SD coupling and $\Yobk = \etaobk = \ln 1/0.01$.}
				\label{fig:lightq-sigmar}
			\end{minipage}
		\end{figure}
		
We perform fits to the HERA reduced cross section data \cite{Aaron:2009aa} using the three BK equations,
\eqref{eq:kcbk}, \eqref{eq:resumbk} and \eqref{eq:trbk}, 
and compare two running coupling schemes and two initial condition rapidity scales. We find that the CGC framework at NLO describes the data very well. In Fig.~\ref{fig:hera-sigmar} one set of fits are shown and one can see that the different BK equations describe the data comparably. We find that even the precise HERA data cannot differentiate between the BK equations or coupling schemes.

In Fig.~\ref{fig:lightq-sigmar} a fit to a light-quark-only dataset generated by interpolation is shown together with the corresponding fit to the HERA data.
One sees that the CGC framework at NLO can fit the light quark data as well. 
We find that the light-quark-only fits systematically prefer larger proton size with slow $x$-evolution, which
we interpret 
as an indication of
a substantial non-perturbative contribution in the light quark production. This would imply that there is a sizeable theoretical uncertainty related to this contribution and thus to these fits.

\begin{figure}[t]
			\begin{minipage}[c]{0.48\textwidth}
				\includegraphics[width=\textwidth]{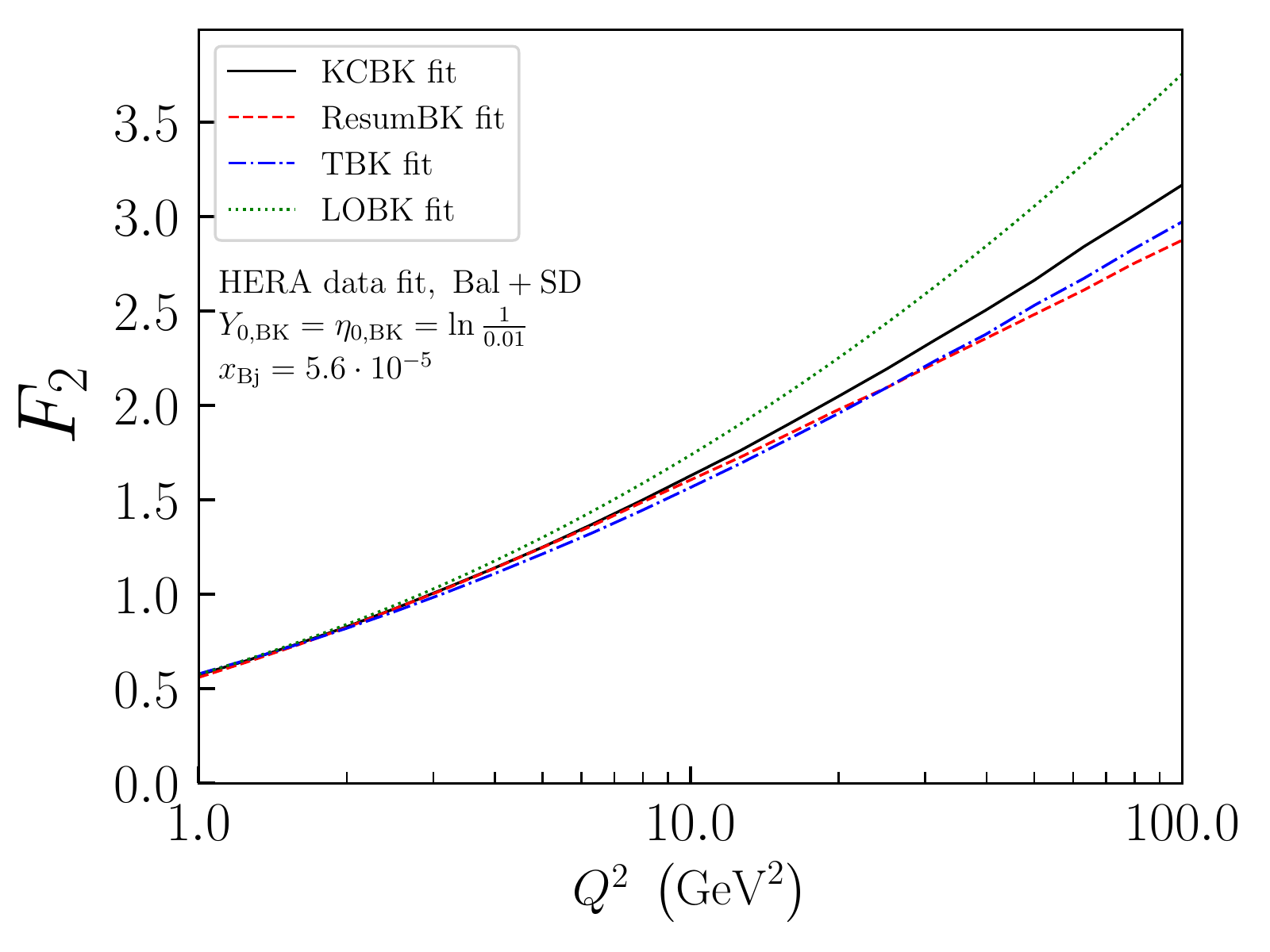}
				\caption{Structure function $F_2$ computed based on HERA data NLO fits with each of the BK equations and a LO fit for comparison at LHeC kinematics.}
				\label{fig:lhec}
			\end{minipage} 
				\hfill
			\begin{minipage}[c]{0.48\textwidth}
				\includegraphics[width=\textwidth]{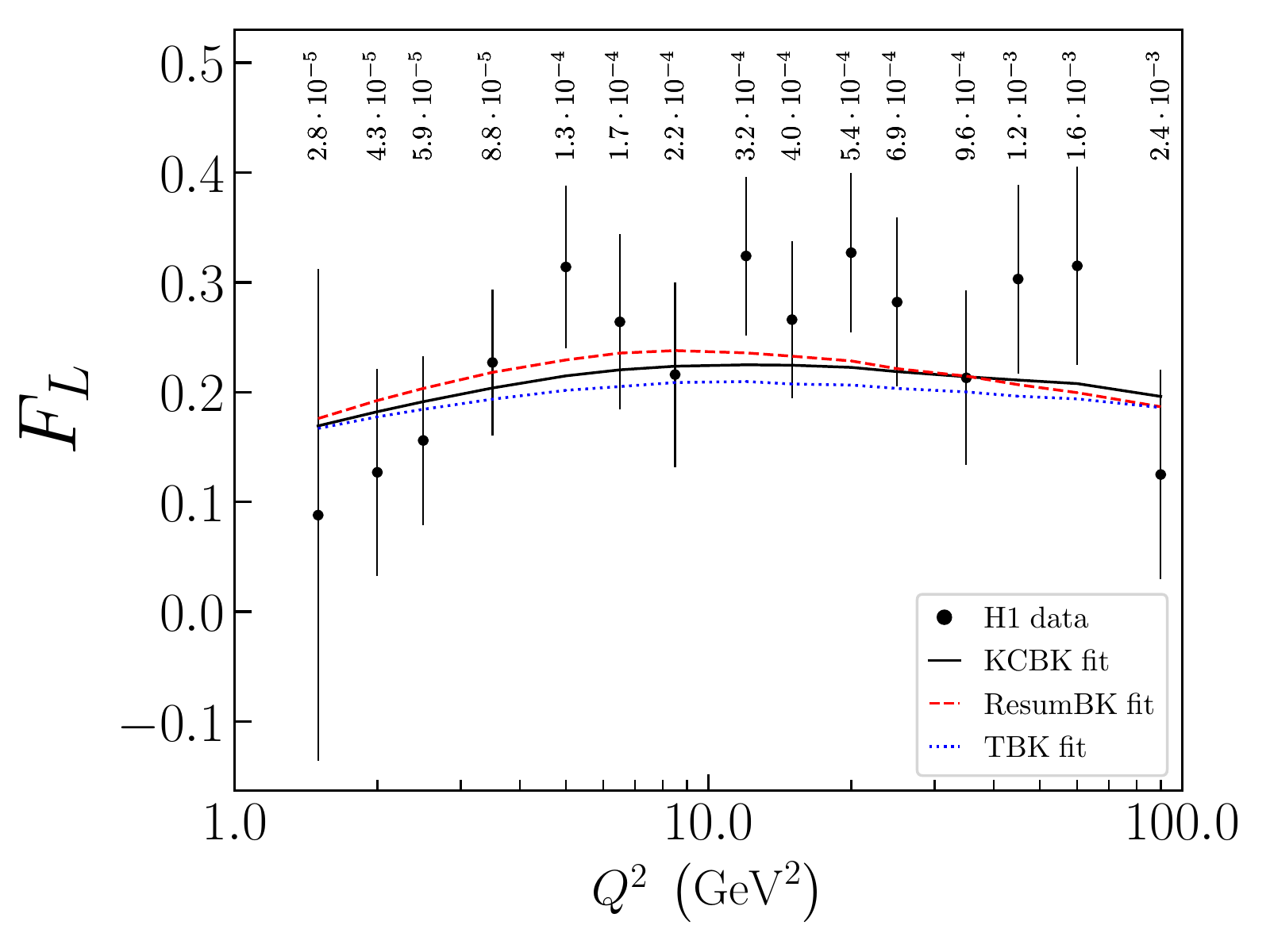}
				\caption{Comparison of H1 measurement of $F_L$~\cite{Andreev:2013vha} and computation using NLO fits with each of the evolution equations. Fits are to HERA data.
				}
				\label{fig:h1fl}
			\end{minipage}
		\end{figure}

Using the fits we extrapolate the structure functions to smaller Bjorken-$x$ into the kinematical range probed by future experiments. The results are shown in Fig.~\ref{fig:lhec}. We find that the NLO corrections are quite stable in the different fit schemes in comparison to the LO result, and that the differences between the schemes are moderate even at LHeC 
kinematics. In Fig.~\ref{fig:h1fl} we compare $F_L$ computed based on the HERA data fits to the H1 collaboration measurement of $F_L$, and we find that the fits describe the data well, and the different schemes are essentially equivalent here.

\section{Conclusions}
We have performed the first fit to HERA reduced cross section data at next to leading order accuracy in the CGC framework using massless quarks.
We use instead approximative versions of the NLOBK equation that resum higher order corrections enhanced by large transverse logarithms. All three BK equations used in this work result in excellent descriptions of the HERA data.

As the DIS structure functions at NLO accuracy are only available for massless quarks, for comparison we generated interpolated dataset for light quark reduced cross section that we used in the fits. 
This generated data is also described excellently as well. Additionally we find systematic changes in the initial condition fit parameters that imply that the light quark cross section contains a sizeable non-perturbative contribution.

\acknowledgments
This work was supported by the Academy of Finland, projects 314764 (H.M) and  321840 (T.L). G.B, H.H and T.L are supported by the European Research Council grant ERC-2015-CoG-681707. Computing resources from CSC – IT Center for Science in Espoo, Finland and from the FGCI (persistent identifier \texttt{urn:nbn:fi:research-infras-2016072533}) were used.

\bibliographystyle{JHEP}
\bibliography{refs}

\providecommand{\href}[2]{#2}\begingroup\raggedright\begin{thebibliography}{10}

\bibitem{Balitsky:1995ub}
I.~Balitsky, \emph{{Operator expansion for high-energy scattering}},
  \href{http://dx.doi.org/10.1016/0550-3213(95)00638-9}{\emph{Nucl. Phys.}
  {\bfseries B463} (1996) 99--160},
  [\href{https://arxiv.org/abs/hep-ph/9509348}{{\ttfamily hep-ph/9509348}}].

\bibitem{Kovchegov:1999yj}
Y.~V. Kovchegov, \emph{{Small x F(2) structure function of a nucleus including
  multiple pomeron exchanges}},
  \href{http://dx.doi.org/10.1103/PhysRevD.60.034008}{\emph{Phys. Rev.}
  {\bfseries D60} (1999) 034008},
  [\href{https://arxiv.org/abs/hep-ph/9901281}{{\ttfamily hep-ph/9901281}}].

\bibitem{Beuf:2020dxl}
G.~Beuf, H.~Hänninen, T.~Lappi and H.~Mäntysaari, \emph{{Color Glass
  Condensate at next-to-leading order meets HERA data}},
  \href{https://arxiv.org/abs/2007.01645}{{\ttfamily 2007.01645}}.

\bibitem{Lappi:2016fmu}
T.~Lappi and H.~Mäntysaari, \emph{{Next-to-leading order Balitsky-Kovchegov
  equation with resummation}},
  \href{http://dx.doi.org/10.1103/PhysRevD.93.094004}{\emph{Phys. Rev.}
  {\bfseries D93} (2016) 094004},
  [\href{https://arxiv.org/abs/1601.06598}{{\ttfamily 1601.06598}}].

\bibitem{Ducloue:2019ezk}
B.~Ducloué, E.~Iancu, A.~H. Mueller, G.~Soyez and D.~N. Triantafyllopoulos,
  \emph{{Non-linear evolution in QCD at high-energy beyond leading order}},
  \href{http://dx.doi.org/10.1007/JHEP04(2019)081}{\emph{JHEP} {\bfseries 04}
  (2019) 081}, [\href{https://arxiv.org/abs/1902.06637}{{\ttfamily
  1902.06637}}].

\bibitem{Beuf:2014uia}
G.~Beuf, \emph{{Improving the kinematics for low-$x$ QCD evolution equations in
  coordinate space}},
  \href{http://dx.doi.org/10.1103/PhysRevD.89.074039}{\emph{Phys. Rev.}
  {\bfseries D89} (2014) 074039},
  [\href{https://arxiv.org/abs/1401.0313}{{\ttfamily 1401.0313}}].

\bibitem{Iancu:2015vea}
E.~Iancu, J.~D. Madrigal, A.~H. Mueller, G.~Soyez and D.~N. Triantafyllopoulos,
  \emph{{Resumming double logarithms in the QCD evolution of color dipoles}},
  \href{http://dx.doi.org/10.1016/j.physletb.2015.03.068}{\emph{Phys. Lett.}
  {\bfseries B744} (2015) 293--302},
  [\href{https://arxiv.org/abs/1502.05642}{{\ttfamily 1502.05642}}].

\bibitem{Iancu:2015joa}
E.~Iancu, J.~D. Madrigal, A.~H. Mueller, G.~Soyez and D.~N. Triantafyllopoulos,
  \emph{{Collinearly-improved BK evolution meets the HERA data}},
  \href{http://dx.doi.org/10.1016/j.physletb.2015.09.071}{\emph{Phys. Lett.}
  {\bfseries B750} (2015) 643--652},
  [\href{https://arxiv.org/abs/1507.03651}{{\ttfamily 1507.03651}}].

\bibitem{Beuf:2016wdz}
G.~Beuf, \emph{{Dipole factorization for DIS at NLO: Loop correction to the
  $\gamma^*_{T,L}\to q\overline q$ light-front wave functions}},
  \href{http://dx.doi.org/10.1103/PhysRevD.94.054016}{\emph{Phys. Rev.}
  {\bfseries D94} (2016) 054016},
  [\href{https://arxiv.org/abs/1606.00777}{{\ttfamily 1606.00777}}].

\bibitem{Beuf:2017bpd}
G.~Beuf, \emph{{Dipole factorization for DIS at NLO: Combining the $q\bar{q}$
  and $q\bar{q}g$ contributions}},
  \href{http://dx.doi.org/10.1103/PhysRevD.96.074033}{\emph{Phys. Rev.}
  {\bfseries D96} (2017) 074033},
  [\href{https://arxiv.org/abs/1708.06557}{{\ttfamily 1708.06557}}].

\bibitem{Hanninen:2017ddy}
H.~Hänninen, T.~Lappi and R.~Paatelainen, \emph{{One-loop corrections to light
  cone wave functions: the dipole picture DIS cross section}},
  \href{http://dx.doi.org/10.1016/j.aop.2018.04.015}{\emph{Annals Phys.}
  {\bfseries 393} (2018) 358--412},
  [\href{https://arxiv.org/abs/1711.08207}{{\ttfamily 1711.08207}}].

\bibitem{Ducloue:2017ftk}
B.~Ducloué, H.~Hänninen, T.~Lappi and Y.~Zhu, \emph{{Deep inelastic
  scattering in the dipole picture at next-to-leading order}},
  \href{http://dx.doi.org/10.1103/PhysRevD.96.094017}{\emph{Phys. Rev.}
  {\bfseries D96} (2017) 094017},
  [\href{https://arxiv.org/abs/1708.07328}{{\ttfamily 1708.07328}}].

\bibitem{Aaron:2009aa}
{\scshape H1, ZEUS} collaboration, F.~D. Aaron et~al., \emph{{Combined
  Measurement and QCD Analysis of the Inclusive $e^\pm p$ Scattering Cross
  Sections at HERA}},
  \href{http://dx.doi.org/10.1007/JHEP01(2010)109}{\emph{JHEP} {\bfseries 01}
  (2010) 109}, [\href{https://arxiv.org/abs/0911.0884}{{\ttfamily 0911.0884}}].

\bibitem{Andreev:2013vha}
{\scshape H1} collaboration, V.~Andreev et~al., \emph{{Measurement of inclusive
  $e p$ cross sections at high $Q^2$ at $\sqrt s =$ 225 and 252 GeV and of the
  longitudinal proton structure function $F_L$ at HERA}},
  \href{http://dx.doi.org/10.1140/epjc/s10052-014-2814-6}{\emph{Eur. Phys. J.
  C} {\bfseries 74} (2014) 2814},
  [\href{https://arxiv.org/abs/1312.4821}{{\ttfamily 1312.4821}}].

\end{thebibliography}\endgroup

\end{document}